\documentclass[12pt]{article}

\usepackage{newtxtext,newtxmath}
\usepackage{graphicx}

\usepackage[letterpaper,margin=1in]{geometry}
\renewenvironment{abstract}
	{\quotation}
	{\endquotation}

\date{}

\makeatletter
\renewcommand{\fnum@figure}{\textbf{Figure \thefigure}}
\renewcommand{\fnum@table}{\textbf{Table \thetable}}
\makeatother

\usepackage{scicite}

\usepackage{url}

\def\scititle{
	Real-Space Plasmon Imaging Reveals Modified Electronic Structure of Gold at the Monolayer Limit 
}
\title{\bfseries \boldmath \scititle}

\author{
	Andrei~Bylinkin$^{1\ast}$,
	Philippe~Roelli$^{1}$,
	Naveen~Shetty$^{2}$,
    Rositsa~Yakimova$^{3}$,\and
    Ulrich~Starke$^{4}$,
    Camilla~Coletti$^{4}$,
    Stiven~Forti$^{5}$,
    Alexei~Zakharov$^{6}$,\and
    Vyacheslav~M.~Silkin$^{7,8,9}$,
    Samuel~Lara-Avila$^{2}$,
    Rainer~Hillenbrand$^{1,9,10\ast}$    
    \and
	\small$^{1}$ CIC nanoGUNE BRTA, Donostia-San Sebastián \& 20018 Spain. \and
	\small$^{2}$ Department of Microtechnology and Nanoscience, Chalmers University of Technology, Gothenburg \and\small \& 412 96, Sweden.\and
    \small$^{3}$ Department of Physics, Chemistry and Biology (IFM), Linköping University, Linköping \& 581 83, Sweden \and
    \small$^{4}$ Max Planck Institute for Solid State Research,  Stuttgart \& 70569, Germany.\and
    \small$^{5}$ Center for Nanotechnology Innovation@NEST, Istituto Italiano di Tecnologia, Pisa \& 56127, Italy.\and
    \small$^{6}$ MAX IV Laboratory, Lund University, Lund \& 22484, Sweden.\and
    \small$^{7}$ Donostia International Physics Center, Donostia-San Sebastián \& 20018, Spain.\and
    \small$^{8}$ Departamento de Polímeros y Materiales Avanzados: Física, Química y Tecnología, \and\small Facultad de Ciencias Químicas, Universidad del País Vasco UPV-EHU, \and\small Donostia-San Sebastián \& 20080, Spain.\and
    \small$^{9}$ IKERBASQUE, Basque Foundation for Science, Bilbao \& 48013, Spain.\and
    \small$^{10}$ Department of Electricity and Electronics, University of the Basque Country (UPV/EHU), \and\small Leioa \& 48940, Spain.\and
	\small$^\ast$Corresponding authors. Emails: a.bylinkin@nanogune.eu, r.hillenbrand@nanogune.eu\and
}

\begin{document} 

% Insert the title and author list
\maketitle

\newpage
% Abstract, in bold
% There are strict length limits, and not all formats have abstracts.
% Consult the journal instructions to authors for details.
% Do not cite any references in the abstract.
\begin{abstract} \bfseries \boldmath

Atomically thin materials exhibit electronic and optical properties distinct from their three dimensional counterparts. For metals, particularly gold, monolayer studies remain largely unexplored due to fabrication and characterisation challenges. Here we report the first optical study of a stable quasi-freestanding gold monolayer formed by Au intercalation between graphene and SiC. Mid-infrared nanoimaging reveals plasmon-polaritons with wavelengths nearly an order of magnitude shorter than free-space light. Analysis of their dispersion using a Drude model yields a relaxation time of 
$\tau = 18\,$fs, comparable to bulk gold, and a Drude weight of 
$D = 1.3\,$mS$\cdot$eV, nearly twice the bulk expectation. These results establish monolayer gold as a two-dimensional metal, opening opportunities for nanoscale photonics, plasmonics and ultra-thin electronics.

\end{abstract}

% The first paragraph of any Science paper does NOT have a heading
% Nor is it indented
\noindent Gold is a remarkable material with exceptional optical\cite{Barnes2003, Dreaden2012}, electronic\cite{Karna2023,Brorson1987,Karaman2024}, and chemical properties\cite{Karnwal2024, Hashmi2006}. When thinned down to nanometre, gold exhibits different behaviour compared to its bulk form\cite{Bowman2024}. Ultrathin gold films show modified band structures\cite{koroteev_quantum-size_2024}, unusual optical responses\cite{Maniyara2019,manjavacas_tunable_2014}, and enhanced nonlinear effects owing to strong confinement of electrons and electromagnetic fields~\cite{rodriguez_echarri_nonlinear_2023,grosmann_nonclassical_2019, Qian2016, Kauranen2012, pan_large_2024}. Theoretical studies further predict that a single-atom-thick gold layer could display extraordinary electrical conductivity~\cite{zhao_electrical_2024} and enhanced nonlinear optical performance~\cite{pan_large_2024}, making it a compelling platform for nanophotonics and quantum materials research. However, fabricating continuous, atomically thin gold layers is extremely challenging because of gold’s high surface energy and strong tendency to dewet. Unlike van der Waals materials, metallic layers cannot be exfoliated because of strong interlayer metallic bonding. Early efforts achieved approximately 57 nm$^{2}$ in size freestanding monolayer Au nanoribbons by dealloying inside transmission electron microscopes~\cite{wang_free-standing_2019}, and recently, chemical synthesis has produced single-atom-thick gold flakes with lateral sizes up to about 100 nm~\cite{kashiwaya_synthesis_2024}. While these approaches demonstrated the feasibility of single-atom-thick gold, they are limited to nanoscale flakes and cannot achieve continuous, large-area layers, which among others has prevented electric transport and optical spectroscopy studies. A promising route to overcome these limitations is the intercalation of Au atoms between silicon carbide (SiC) and monolayer graphene~\cite{forti2020semiconductor}, which  provides a scalable route to form a $\mu$m-size area of atomically thin gold layers.

In the case of Au intercalated between SiC and graphene, the first intercalated Au layer forms a two-dimensional Au(111)-like plane that serves as a termination layer for the SiC substrate. This layer is hereafter referred to as zero-layer gold (ZL-Au) (Fig.~\ref{fig:1}A), by analogy with zero-layer graphene (carbon buffer layer) in graphene/SiC systems~\cite{emtsev2008sic,starke2009epitaxial_graphene,forti2014epitaxial_graphene}. Unlike zero-layer graphene, which lacks clear band dispersion, ZL-Au exhibits well-defined dispersive bands, despite each Au atom being bonded to a Si atom at the SiC interface. Angle-resolved photoemission spectroscopy (ARPES) shows that these bands lie below the Fermi level without crossing it, establishing the semiconducting character of ZL-Au, and further shows that the graphene layer above is strongly \textit{n}-type doped (Fermi level $\approx 700$ meV)~\cite{forti2020semiconductor}. In certain regions, for specific intercalation parameters, a second Au layer grows in a controllable manner atop the first, forming a two-layer system with graphene on top. ARPES shows that these two-layer regions are metallic, with states crossing the Fermi level, and induce weak \textit{p}-type doping in graphene (Fermi level $\approx 150$ meV)~\cite{forti2020semiconductor}. The metallic behavior is primarily associated with the second Au layer, which is largely decoupled from the substrate due to the presence of the underlying ZL-Au and can be considered a quasi-freestanding monolayer (ML-Au). Consistently, X-ray photoelectron spectroscopy (XPS) studies of ML-Au show that the Au 4f core levels remain shifted relative to bulk Au and are split into two components corresponding to ZL-Au (Au–Si–bonded) and ML-Au, rather than merging into a single bulk Au 4f peak~\cite{forti2020semiconductor}. These well-defined heterostructures provide a versatile platform for investigating the electronic and optical properties of gold monolayer. To date, neither optical nor electrical studies have been performed, and experimental insights into the carrier dynamics and the ability of ML-Au to sustain plasmon polaritons remain largely unexplored.

Here, we study the mid-infrared conductivity of a stable ML-Au, formed by lateral gold intercalation between graphene and SiC~\cite{Kim2020ambipolar}. Using scattering-type scanning near-field optical microscopy (s-SNOM)\cite{hillenbrand2025nanoscopy} operating between 1500 cm$^{-1}$ and 1900 cm$^{-1}$, we find that ML-Au exhibits stronger near-field reflectivity than ZL-Au, confirming the metallic character of the ML-Au. Importantly, we observe plasmon polaritons in ML-Au and employ polariton interferometry to determine their dispersion relation, providing direct evidence of propagating collective charge excitations in the monolayer. Fitting the dispersion with a Drude conductivity model yields the Drude weight ($D \propto n/m_{\text{eff}}$) and the electron relaxation time $\tau$. We obtain $\tau = 18$ fs, comparable to bulk gold, while $D = 1.3$ mS·eV is nearly twice the bulk expectation but in good agreement with density functional theory (DFT) for an isolated Au(111) monolayer. Using this approach, we can directly probe the optical conductivity of ML-Au at the atomic-layer limit. The results further establish gold monolayers as a platform for investigating charge dynamics in atomically thin materials and for potential applications in ultra-thin conductors, nanoscale waveguides, and hybrid nanophotonic systems.

\subsection*{Results}

\begin{figure}
\centering
\includegraphics[scale=1.1]{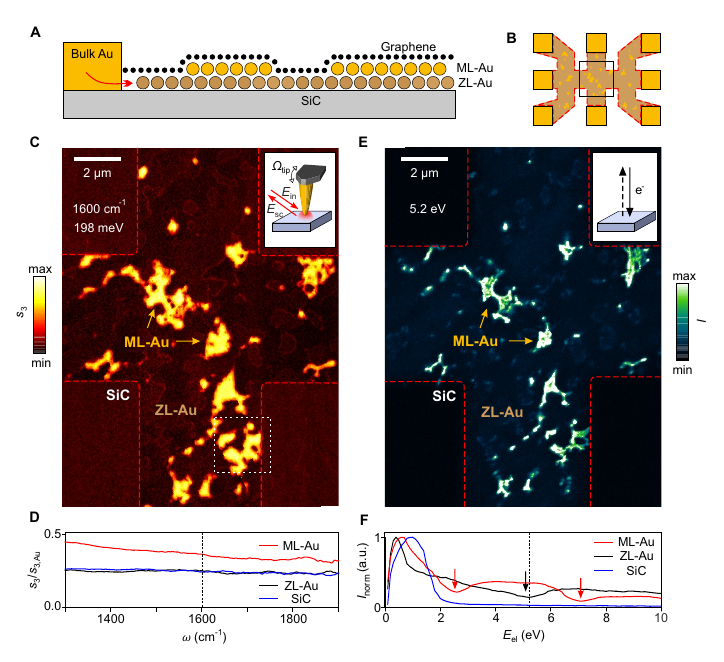 }
  \caption{ \textbf{s-SNOM and LEEM}.
  (\textbf{A}) Illustration of the zero-layer (ZL) and quasi-freestanding monolayer (ML) Au regions located between the SiC(4H) substrate and graphene. Red arrow indicates the intercalation of Au atoms.
  (\textbf{B}) Illustration of the Hall bar structure of the intercalated area. The black square indicates the scanned region.
  (\textbf{C}) s-SNOM amplitude, $s_3$, image at $\omega$ = 1600~cm$^{-1}$. The inset in the top right corner is a schematic of the s-SNOM experiment, where $E_{\text{in}}$ and $E_{\text{sc}}$ denote the incident and tip-scattered electric fields. The white dashed square marks the zoomed-in area shown in Fig.~\ref{fig:2}A,B.
  (\textbf{D}) Representative nano-FTIR spectra, $s_3/s_{\text{3,Au}}$, of ZL-Au, ML-Au, and SiC.
  (\textbf{E}) LEEM image at $E_{\text{el}}$= 5.2~eV. The inset in the top right corner is a schematic of the LEEM experiment. Solid and dashed black lines indicate the incident and reflected electrons, respectively.   
  (\textbf{F}) LEEM-IV spectra of ZL-Au, ML-Au, and SiC. Black and red arrows mark the dips in spectra.  
  (\textbf{C,E}) Red dashed lines mark the boundary between the bare SiC regions and the region with intercalated Au. }
  \label{fig:1}
\end{figure}

Fig.~\ref{fig:1}A shows the cross section of the Au-intercalated graphene/SiC heterostructure. The gold-colored rectangular on the left side illustrates the bulk Au contact on SiC substrate. The red arrow indicates the lateral intercalation process of Au atoms between graphene and the SiC substrate~\cite{Kim2020ambipolar} (see Methods for the details). The intercalation process begins with the formation of ZL-Au (brown dots), after which ML-Au (gold dots) typically appears as discontinuous islands on top.

To study the optical properties of the sample, we perform nanoimaging and nanoscale Fourier transform IR spectroscopy (nano-FTIR) using a s-SNOM~\cite{hillenbrand2025nanoscopy} (see Methods). In s-SNOM a metallized atomic force microscope (AFM) tip acts as a near-field probe (illustrated in Fig.~\ref{fig:1}C). The tip, oscillating at frequency $\mathit{\Omega}_{\text{tip}} \approx 300$  kHz, is illuminated by incident radiation (indicated by the red arrow, $E_{\text{in}}$). Via lightning-rod effect, the tip concentrates the radiation into a nanoscale near-field spot at the tip apex. Interferometric detection and demodulation at a frequency $3\mathit{\Omega}_{\text{tip}}$ of the tip-scattered field yields complex-valued nanoscale-resolved images or spectra, expressed by amplitude and phase, $s_3$ and $\varphi_3$, respectively~\cite{ocelic2006pseudo}(see Methods). Fig.~\ref{fig:1}C shows an amplitude image $s_3$ of one cross of a Hall bar device (Methods) recorded with monochromatic illumination at 1600~cm$^{-1}$. The scanned region is marked by a black square in Fig.~\ref{fig:1}B. The cross-shaped intercalated area (outlined by a red dashed line) lies on top of the SiC, while surrounding regions remain bare SiC. Within the intercalated area, bright ML-Au islands are observed, separated by darker ZL-Au regions. Representative nano-FTIR spectra for the two regions (Fig.~\ref{fig:1}D) reveal that the near-field reflectivity of ML-Au (red curve) exceeds that of ZL-Au (black curve) by an approximate factor of $1.5$ at $1800\ \mathrm{cm}^{-1}$ and $2$ at $1200\ \mathrm{cm}^{-1}$, demonstrating the much stronger near-field response of the ML-Au layer. In contrast, the response of ZL-Au is nearly indistinguishable from that of bare SiC (blue curve), consistent with the very weak near-field response expected for an atomically thin semiconductor.

To corroborate the layer assignment, we perform low-energy electron microscopy (LEEM) imaging and spectroscopy (Fig.~\ref{fig:1}E; see Methods). LEEM provides real-space imaging with nanometre-scale resolution and reveals local electronic and structural properties through intensity–voltage (IV) spectroscopy. Fig.~\ref{fig:1}E shows a LEEM image recorded at electron energy $E_{\text{el}} = 5.2$ eV. Fig.~\ref{fig:1}F shows LEEM-IV spectra from ZL-Au regions and ML-Au islands (black and red curves, respectively), and, for comparison, from the bare SiC substrate (blue curve), which shows no characteristic features. The spectrum from ZL-Au exhibits a single dip (black arrow) originating from the graphene layer, whereas the ML-Au spectrum exhibits two dips (red arrows), corresponding to one graphene and one ML-Au layers. No extra dip appears for ZL-Au, as each Au atom bonds to an Si atom at the SiC interface, preventing the emergence of a separate electronic state in the LEEM-IV spectrum~\cite{hibino2008microscopic}. These assignments are confirmed through comparison with LEEM-IV spectra from well-characterized Au-intercalated samples~\cite{forti2020semiconductor}.

\begin{figure}
\centering
\includegraphics[scale=1.1]{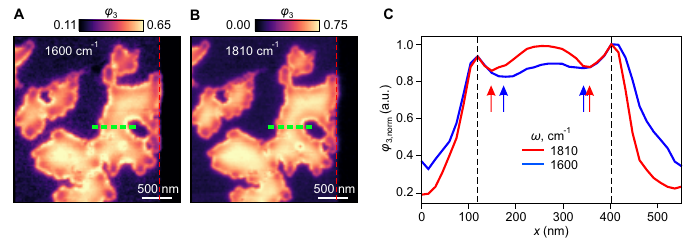}
  \caption{\textbf{Near-field phase images}. 
  (\textbf{A,B})  Near-field phase images, $\varphi_3$, of the region marked by the white dashed square in Fig.~\ref{fig:1}c at frequencies $\omega$ = 1600 cm$^{-1}$ and 1810 cm$^{-1}$, respectively. The vertical red dashed line marks the boundary of the intercalated region. The horizontal green dashed line indicates the position along which the line profiles in Fig.~\ref{fig:2}C were extracted.
  (\textbf{C}) Blue and red curves show the phase line profiles extracted along the horizontal green dashed line, shown in panels A and B, respectively. The vertical dashed lines mark the maxima in both of the line profiles. The blue and red arrows mark the local minima in the line profile at $\omega$ = 1600 cm$^{-1}$ and 1810 cm$^{-1}$, respectively.  
  }
  \label{fig:2}
\end{figure}

Interestingly, the bright ML-Au islands appear spatially inhomogeneous, with blurred edges, resembling a plasmon-polariton response. To investigate this, we zoom in on the island marked by the dashed white square in Fig.~\ref{fig:1}D. Fig.~\ref{fig:2}A,B show phase images, $\varphi_3$, of this island at two illumination frequencies (phase images are shown because they better reveal polaritonic effects~\cite{jing2023phaseresolved}). In both images, maxima and minima appear parallel to the island edges, reminiscent of the interference patterns generated by tip-launched polaritons~\cite{Chen2022topological,dai2019monolayerhbn,jing2023phaseresolved}. This interpretation is further supported by the line profiles shown in Fig.~\ref{fig:2}C, extracted along the dashed green lines in Fig.~\ref{fig:2}A,B. The fringe minima (marked by arrows) shift toward the island edge with increasing frequency, a characteristic behaviour of propagating polaritons with positive phase velocity. These results provide strong experimental evidence of the existence of plasmon-polariton modes on the ML-Au islands.

\begin{figure}
\centering
\includegraphics[scale=1]{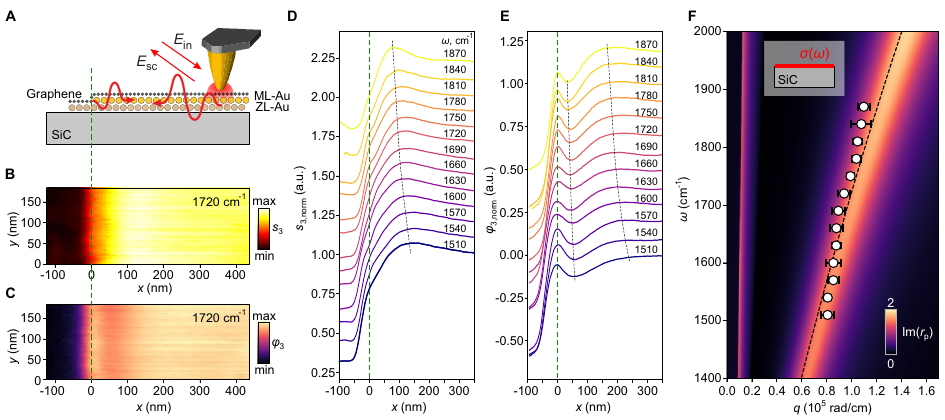}
  \caption{\textbf{Polariton interferometry experiment}. 
  (\textbf{A}) Illustration of the nanoimaging experiment. $E_{\text{in}}$ and $E_{\text{sc}}$ denote the incident and tip-scattered electric fields. Red decaying sine waves illustrate plasmon polaritons propagating in the ML-Au region. 
  (\textbf{B,C}) Near-field amplitude, $s_3$, and phase, $\varphi_3$, images at a frequency of $\omega$ = 1720 cm$^{-1}$, respectively. 
  (\textbf{A,B,C}) The vertical green dashed line marks the edge of the ML-Au region. 
  (\textbf{D,E}) Near-field amplitude, $s_{3,\text{norm}}$, and phase line profiles,  $\varphi_{3,\text{norm}}$, of ML-Au region at different frequencies, recorded perpendicular to the edge. Thin dashed lines indicate the positions of the maxima or minima.
  (\textbf{F}) White circles show the real part of the plasmon-polariton momentum obtained by complex-valued fitting of line profiles in panels D and E. The black dashed curve shows the calculated plasmon-polariton dispersion for the model system, which consists of a sheet characterized by a 2D Drude-type optical conductivity $\sigma(\omega)$ on an SiC substrate, illustrated in the inset. The parameters of the Drude conductivity were determined by fitting the calculated dispersion to experimental data points (white circles). The colour plot shows the calculated imaginary part of the Fresnel reflection coefficient, Im$(r_{\text{p}})$, for the same system.
  }
  \label{fig:3}
\end{figure}

To verify the presence of polaritons in ML-Au and investigate their origin, we perform a systematic polariton interferometry experiment~\cite{hillenbrand2025nanoscopy,chen2012graphene_plasmons,fei2012graphene_plasmons}  on one ML-Au island at multiple frequencies, allowing us to extract the dispersion of these polaritons. Fig.~\ref{fig:3}A illustrates the polariton interferometry experiment. The near field at the tip apex launches polaritons, which propagate to the ML-Au island's edge (marked by a vertical green line), reflect and return to the tip, where the polaritons' field interferes with the near field at the tip apex, leading to the spatial variation of the tip-scattered field $E_{\text{sc}}$. To avoid additional reflections from other edges or defects, we select a micrometre-sized ML-Au island with a well-defined edge. Fig.~\ref{fig:3}B,C show exemplary amplitude and phase images across the edge of the ML-Au island at $\omega = 1720 $ cm$^{-1}$. We extract amplitude and phase line profiles across the edge from images acquired at 13 different frequencies spanning 1510–1870~cm$^{-1}$ (Fig.~\ref{fig:3}D,E). These line profile reveal amplitude maxima accompanied by alternating phase minima and maxima (indicated by black dashed lines). The corresponding complex near-field line profiles, $s_3(x)e^{i\varphi_3(x)}$, trace spiral paths, which can be fitted to radially propagating damped waves~\cite{Chen2022topological} (Methods), confirming the presence of tip-launched damped polaritons. Fitting these profiles yields the polariton complex-valued momentum $q_{\text{p}} + i\kappa_{\text{p}}$, with the real part, $q_{\text{p}}$, plotted versus $\omega$ in Fig.~\ref{fig:3}F to reveal the polariton dispersion. This dispersion confirms that ML-Au supports propagating plasmon polaritons and, importantly, provides a quantitative basis for analyzing its underlying optical conductivity, $\sigma(\omega)$, because of the direct connection between polariton momentum and $\sigma$~(Eq.~\ref{dispersion}). 

We highlight the remarkable eightfold compression of the plasmon wavelength, $\lambda_p = 2\pi/ q_p < 8\lambda_0 $, compared to the free-space wavelength $\lambda_0 $ at the same frequency. At first glance, this may seem surprising for Au plasmon polaritons in the mid-infrared spectral range. However, calculations of the plasmon-polariton dispersion for Au layers with thicknesses between 1 and 10 nm show that such compression arises naturally when the thickness is reduced well below 5 nm.

In the regime where the polariton momentum greatly exceeds the photon momentum in free space ($q_{\mathrm{p}} \gg \omega/c$), the polariton dispersion of a 2D conductive sheet on a substrate (inset Fig.~\ref{fig:3}F) is generally determined by the sheet conductivity, $\sigma(\omega)$, and the dielectric function of the substrate, $\varepsilon_{\text{s}}$. In this limit, the polariton dispersion is given by~\cite{jablan2009plasmonics}:
\begin{equation}
\label{dispersion}
    q_{\text{p}}(\omega)+i\kappa_{\text{p}}(\omega) = \varepsilon_0 \frac{1+\varepsilon_{\text{s}}}{2}\frac{2i\omega}{\sigma(\omega)},
\end{equation}  
where, in our system, $\varepsilon_{\text{s}} = \varepsilon_{\text{SiC}}$ corresponds to the dielectric function of the SiC substrate~\cite{tiwald1999sic}. 

\begin{figure}
\centering
\includegraphics[scale=1]{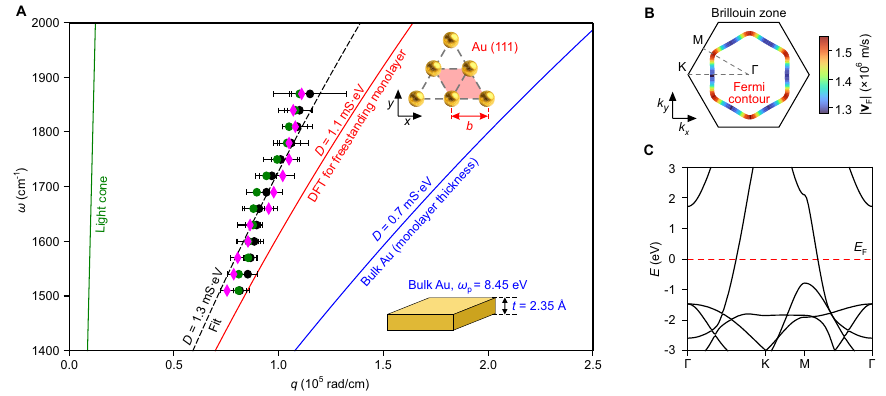}
  \caption{ \textbf{Plasmon-polariton dispersion}.  
  (\textbf{A})  The green and black symbols show the plasmon polariton dispersion obtained by fitting line profiles measured at the same ML-Au island but different days, line profiles are shown in Fig.~\ref{fig:3}D,E. Magenta symbols show the plasmon dispersion obtained by fitting line profiles measured at another ML-Au island. The dashed black line shows the fit of the experimental dispersion, the same line as in Fig.~\ref{fig:3}F. The green curve shows the photon dispersion. The blue curve shows the plasmon–polariton dispersion obtained using the Drude weight derived for a thin Au film with a finite thickness of $t = 2.35$~\AA{}, as sketched in the bottom inset, and employing the bulk dielectric function~\cite{olmon2012optical}. The red curve shows the plasmon-polariton dispersion obtained using the Drude weight derived from the DFT-calculated band structure of isolated freestanding monolayer Au (111) (shown in panel C). An illustration of freestanding Au monolayer (111) with the unit cell marked by red area shown as inset next to the red curve. 
 (\textbf{B}) The Fermi contour of DFT-calculated freestanding monolayer Au (111) with lattice constant $b = 2.88$~\AA{}, with colour indicating the magnitude of the Fermi velocity $|\mathbf{v}_{\text{F}}(\mathbf{k})|$ at each point on the Fermi contour. The black hexagon represents the first Brillouin zone. 
 (\textbf{C}) DFT-calculated electronic band structure of freestanding monolayer Au (111). The dashed red line marks the Fermi level.}
  \label{fig:4}
\end{figure}

To relate the polariton dispersion to the carrier properties of ML-Au, we model the sheet conductivity using the Drude model (Eq.~\ref{conductivity}). In analogy with the standard Drude description of bulk Au in the infrared range~\cite{olmon2012optical}, we neglect interband contributions, as further supported by the DFT electronic band structure shown in Fig.~\ref{fig:4}C. While more advanced quantum and non-local models exist, these corrections typically introduce only minor changes to the frequency dependence of $q_{\text{p}}(\omega)$~\cite{echarri_quantum_2019,alonsoGonzalez2016acoustic}. Consequently, the Drude model provides a fundamental and physically intuitive framework for interpreting our experimental results. The Drude conductivity is given by:
\begin{equation}
\label{conductivity}
    \sigma(\omega) = \frac{i}{\pi}\frac{D}{\omega+i\tau^{-1}},
\end{equation}
where $D$ is the 2D Drude weight (hereinafter referred to as Drude weight) and $\tau$ is the carrier relaxation time. $D$ is a ground-state property that corresponds to the inertia of the many-electron system in the adiabatic limit and is sometimes referred to as the adiabatic charge stiffness \cite{resta2018drude}. In classical physics, the Drude weight is given by $D = \pi e^2 (n_{\text{2D}}/m_{\text{eff}})$~\cite{ashcroft_solid_1976}, where $n_{\text{2D}}$ is the effective 2D carrier density and $m_{\text{eff}}$ the effective mass.  

Fitting the experimental dispersion (white circles in Fig.~\ref{fig:3}F) with this model shows excellent agreement, demonstrating that the ML-Au conductivity is well described by a Drude response. We obtain a Drude weight of $D = 1.3$~mS$\cdot$eV and a relaxation time of $\tau = 18$~fs (black dashed line in Fig.~\ref{fig:3}F). We note that this relaxation time is close to reported bulk value of $\tau_{\text{bulk}} = 14 \pm 3$~fs~\cite{olmon2012optical}. Repeating the measurements on the same island on a different day, and on another ML-Au island (green and magenta symbols in Fig.~\ref{fig:4}A, respectively), confirms the robustness of the dispersion, particularly the real part of polariton momentum, and extracted Drude weight. We further note that the contribution of graphene to the optical conductivity is negligible, since previous ARPES study~\cite{forti2020semiconductor} of graphene above ML-Au shows a Fermi level of $\mu_{\text{gr}} \approx 150$~meV, corresponding to a Drude weight of $D_{\text{gr}} \approx e^2 \mu_{\text{gr}}/\hbar^2 \approx 0.04~\text{mS·eV}$~\cite{falkovsky2008optical}, which is only about 3\% of the Drude weight extracted from the polariton dispersion.

To interpret the extracted Drude weight of ML-Au, we compare the measured polariton dispersion with two models, one using bulk Au properties and the other using DFT-calculated freestanding monolayer Au with the bulk in-plane lattice constant. In the first model, we use the Drude dielectric function of gold with bulk plasma frequency $\omega_{\mathrm{p}} = 8.45$~eV~ and $\varepsilon_\infty = 1$~\cite{olmon2012optical}, and treat the layer as a thin Au film of thickness $t = 2.35$~\AA{}, corresponding to the spacing between the (111) planes of bulk Au. This yields a Drude weight of $D = \varepsilon_0 \pi \omega_{\mathrm{p}}^2 t = 0.7$~mS$\cdot$eV (blue curve in Fig.~\ref{fig:4}A), much smaller than the experimental value of $D = 1.3$~mS$\cdot$eV. The discrepancy could, in principle, be interpreted as nearly a two-fold increase in effective thickness, corresponding to two conducting bulk-like Au layers rather than the single layer assumed in our calculation. However, we discard this scenario since XPS, ARPES, and LEEM–IV measurements show that the bottom Au layer is bound to the SiC substrate and exhibits semiconducting behavior, as discussed in the Introduction and description of Fig.~1f.

In the second model, we derive the Drude weight from the DFT-calculated electronic band structure of a freestanding Au(111) monolayer (Fig.~\ref{fig:4}C), assuming the bulk lattice constant of $b = 2.88$~\AA{}, which differs by less than 2\% from the lattice constant measured experimentally for ML-Au~\cite{forti2020semiconductor}. The DFT calculation reveals a single band crossing the Fermi level (horizontal dashed red line in Fig.~\ref{fig:4}C), with no vertical interband transitions across the Fermi level having transition energies below $1$~eV, justifying a purely Drude (intraband) description of the conductivity in the mid-IR frequency range. From the band structure, we extract the direction-dependent Fermi momentum, which describes a quasi-hexagonal Fermi contour (FC, colored contour in Fig.~\ref{fig:4}B) in reciprocal space, with the colour indicating the magnitude of the Fermi velocity, $\mathbf{v}_{\text{F}}(\mathbf{k}) =  \frac{1}{\hbar}\nabla_{\mathbf{k}}E(\mathbf{k})$, at each point along the FC. From this band structure, we evaluate the Drude weight tensor $D_{\alpha\beta}$ ($\alpha,\beta=x,y$) using the general semi-classical Brillouin-zone integral \cite{resta2018drude}:
\begin{equation}
 D_{\alpha\beta} = - 2\pi e^{2} \int_{\mathrm{BZ}}\frac{d^{2}k}{(2\pi)^{2}}\;
    f'(E_{\mathbf{k}})\,
    v_{\text{F},\alpha}(\mathbf{k})\,v_{\text{F},\beta}(\mathbf{k})
    \qquad \text{with} \qquad
    v_{\text{F},\alpha}(\mathbf{k})
    = \frac{1}{\hbar}\,\frac{\partial E(\mathbf{k})}{\partial k_{\alpha}},
    \label{eq:Drude_general}
\end{equation}
where $f'(E) =\frac{df(E)}{dE}$ is the derivative of the Fermi–Dirac distribution. The off-diagonal elements of the Drude weight vanish due to the mirror symmetry of the FC, while the six-fold rotational symmetry of the FC yields equal diagonal elements, $D_{xx} = D_{yy}$, so that the Drude weight can be fully described by a single scalar parameter $D = D_{xx} = D_{yy}$. We obtain $D = 1.1$~mS$\,\cdot\,$eV, which is approximately 60\% larger than the Drude weight calculated with the first model assuming bulk Au properties, yet still smaller than the experimental value, $D = 1.3~\mathrm{mS\cdot eV}$. Consequently, the corresponding plasmon-polariton dispersion (red curve in Fig.~\ref{fig:4}A) remains shifted relative to the experimental dispersion (symbols in  Fig.~\ref{fig:4}A). 

The remaining discrepancy between the experimental and theoretical Drude weight discrepancy may arise from charge transfer within the SiC/ZL-Au/ML-Au/graphene heterostructure. We find that the experimental Drude weight can be obtained theoretically by increasing the carrier concentration by approximately 20\%. However, reliably quantifying the charge transfer would require DFT calculations that include the full SiC/ZL-Au/ML-Au/graphene heterostructure with structural relaxation. This represents a substantial challenge and goes beyond the scope of the present study. Overall, despite the remaining small quantitative discrepancies, comparison of the experimentally measured Drude weight with the bulk-Drude model of the Au layer reveals a substantial enhancement, reflecting the distinct electronic structure at the atomic-layer limit, as corroborated by the DFT calculations.

Finally, to gain an intuitive, classical interpretation of the enhanced intraband response, we express the DFT-derived Drude weight in the classical form, $D = \pi e^2 (n_{\mathrm{2D}}/m_{\mathrm{eff}})$, which provides a convenient framework for assessing the roles of carrier density and effective mass. The two dimensional carrier density is extracted directly from the DFT band structure as the area enclosed by the Fermi contour, in accordance with the Luttinger theorem~\cite{ashcroft_solid_1976}, yielding $n_{\mathrm{2D}}^{\mathrm{DFT}} = 1.43 \times 10^{15}\,\mathrm{cm}^{-2}$. Independently, the carrier density can be estimated from the lattice geometry. The rhombic primitive unit cell shown in Fig.~\ref{fig:4}A (red area in the inset of illustration of monolayer Au) contains one Au atom. Assuming one conduction electron per Au atom, the corresponding  electron density derived from the lattice geometry is $n_{\mathrm{2D}}^{\mathrm{geom}} = 1/A$, with the unit cell area  $A = (\sqrt{3}/2)\, b^2$, which gives $n_{\mathrm{2D}}^{\mathrm{geom}} = 2/(\sqrt{3}\, b^2) = 1.39 \times 10^{15}\,\mathrm{cm}^{-2}$. The close agreement between $n_{\mathrm{2D}}^{\mathrm{DFT}}$ and $n_{\mathrm{2D}}^{\mathrm{geom}}$ shows that the carrier density alone cannot account for the enhancement of the Drude weight. This implies that the enhancement originates from a reduced effective mass $m_{\mathrm{eff}}$ compared to the free-electron mass $m_0$. Using the DFT-derived Drude weight together with the corresponding $n_{\mathrm{2D}}^{\mathrm{DFT}}$, we obtain a Fermi-contour-averaged effective mass of $m_{\mathrm{eff}}^{\mathrm{DFT}} \approx 0.8\,m_0$. This value is smaller than the effective mass typically assumed to model monolayer gold, $m_{\mathrm{eff}} \approx m_0$~\cite{manjavacas_tunable_2014,echarri_quantum_2019}, highlighting the impact of strong electron confinement on the electronic transport properties of monolayer Au. Intuitively, in atomically thin Au layers, reduced electronic screening modifies the band structure, which lowers the effective mass and enhances the collective electronic response.

\subsection*{Conclusion}
In summary, our results show that ML-Au formed by intercalation between SiC and graphene behaves as a two-dimensional metal that supports mid-infrared plasmon polaritons. Nanoimaging and nano-FTIR spectroscopy reveal a strongly enhanced near-field response of ML-Au compared to both ZL-Au and the underlying SiC, indicating its metallic character. By performing systematic polariton interferometry on isolated ML-Au islands, we extract the plasmon-polariton dispersion and find that it can be well described by a Drude-type optical conductivity, with a relaxation time comparable to bulk gold and a Drude weight nearly twice the bulk expectation. This pronounced increase in Drude weight is consistent with DFT calculations for a freestanding Au(111) monolayer, indicating that the carrier dynamics of ML-Au are governed by its monolayer electronic structure rather than by bulk-like electronic properties. Altogether, these observations provide direct experimental evidence for the large optical conductivity and polaritonic response of ML-Au layer, holding potential for applications in future photonic and electronic nanodevices.

%Formatting:
%Names of software packages should be set in small capitals e.g.~\textsc{NumPy}.
%Use a non-breaking space between a number and its unit e.g.~7.4~km,
%and thin spaces between different parts of a unit e.g.~12~m\,s$^{-1}$.
%Use $\pm$ (not parentheses) to indicate uncertainties e.g.~$g=9.8\pm0.2$~m\,s$^{-2}$.

% Research Articles and Reviews split the text into sections using headings
% Use a short (up 6 words) descriptive phrase, not generic 'Results' or 'Conclusions'
% Most other formats do not have headings, see the journal instructions to authors for details

%\subsection*{An example heading}
%Research Articles and Reviews use headings to split the main text into sections;most other formats do not have headings.

%Length limits:
%The \textit{Science}-family journals impose limits on the number of words, figures\slash
%tables, and references cited in the main text. The limits vary between the journals and article types. Refer to the instructions to authors on the journal website for the current limits.

% If your text is very short you might need to uncomment the following line to avoid
% layout problems with the figures and tables.
\newpage

%%%%%%%%%%%%%%%% MAIN TEXT FIGURES %%%%%%%%%%%%%%%

%%%%%%%%%%%%%%%% REFERENCES %%%%%%%%%%%%%%%

\clearpage % Clear all remaining figures and tables then start a new page

% The list of references goes after the main text and before the acknowledgements
% When preparing an initial submission, we recommend you use BibTeX, like this:
%
\bibliography{main_biblio} % for a file named science_template.bib
\bibliographystyle{sciencemag}

% After the paper has completed peer review and been revised ready for acceptance,
% you should comment out the lines above and copy-paste the contents of your .bbl
% file here instead. This will help ensure that our conversion software works correctly.
% Remember to re-run BibTeX first - check the timestamp!
%
% Example of the first three entries copy-pasted from science_template.bbl:
%
%\begin{thebibliography}{1}
%
%\bibitem{example}
%A.~N. {Author}, An example reference. \emph{Journal of Improbable Research}
%  \textbf{1}, 67 (2020).
%
%\bibitem{example2}
%F.~M. {Surname}, S.~{Author}, A second example. \emph{Interesting Research
%  Letters} \textbf{32}, 897 (2019).
%
%\bibitem{example_preprint}
%P.~{One}, P.~{Two}, P.~{Three}, {An unpublished preprint}. \emph{preprint}
%  (2021), arXiv:2101.12345.
%
%\end{thebibliography}

%%%%%%%%%%%%%%%% ACKNOWLEDGEMENTS %%%%%%%%%%%%%%%
\newpage
\section*{Acknowledgments}
The authors thank Alejandro Manjavacas, Andrey Borisov, Daniel Hernangomez and Emilio Artacho for the fruitful discussion.

We also thank MAX-lab (Lund, Sweden) for the allocation of synchrotron radiation beamtime. Support by the staff at the I311 beamline and the MAXPEEM instrument is gratefully acknowledged.

\paragraph*{Funding:}
The work received financial support from grant CEX2020-001038-M and grant PID2024-156602NB-I00 (CEDANO), funded by MICIU/AEI/10.13039/501100011033 and by ERDF/EU. This work was jointly supported by Chalmers Area of Advance Nano, Chalmers Area
of Advance Materials, 2D-Tech VINNOVA competence Center(Ref.2024-03852), and the Swedish Research Council VR under Contract No. 2021-05252.

P.R. acknowledges financial support from the Swiss National Science Foundation (Grant No. 206926) and from the European Union's Horizon 2020 research and innovation programme under the Marie Sklodowska-Curie grant agreement No. 101065661. V.M.S. acknowledges financial support by Grant PID2022-139230NB-I00 funded by MCIN/AEI/10.13039/501100011033. S.F. and U.S. were supported by the Deutsche Forschungsgemeinschaft in the framework of the Priority Program 1459 Graphene (Sta315/8-2).

\paragraph*{Competing interests:}
R.H. is a co-founder of Neaspec GmbH, which now is a part of attocube systems GmbH, a company producing s-SNOM systems, 
such as the one used in this study. The remaining authors declare no competing interests.

%%%%%%%%%%%%%%%% END OF MAIN TEXT %%%%%%%%%%%%%%%

\end{document}